\documentclass[aps, prl, superscriptaddress, twocolumn, letterpaper, 10pt, amsfonts, amsmath, amssymb]{revtex4-2}
\usepackage{graphicx}% Include figure files
\usepackage[hidelinks]{hyperref}% add hypertext capabilities
\usepackage{physics}
\usepackage{float}
\usepackage{xcolor}
\usepackage{soul}

\definecolor{ABcolor}{cmyk}{1,0,0,0.6}

\begin{document}
\title{Boundary effects in two-band superconductors}

\date{\today}
\author{Andrea Benfenati} \email{alben@kth.se}
\affiliation{Department of Physics, The Royal Institute of Technology, Stockholm SE-10691, Sweden}
\author {Albert Samoilenka}
\affiliation{Department of Physics, The Royal Institute of Technology, Stockholm SE-10691, Sweden}
\author {Egor Babaev}
\affiliation{Department of Physics, The Royal Institute of Technology, Stockholm SE-10691, Sweden}

\begin{abstract}
We present a microscopic study of the behavior of the order parameters near boundaries of a two-band superconducting material, described by the standard tight-binding Bardeen-Cooper-Schrieffer model. We find superconducting surface states. The relative difference between bulk and surface critical temperatures is a nontrivial function of the interband coupling strength. 
For superconductors with weak interband coupling, boundaries induce variations of the gaps with the presence of multiple length scales,  despite non-zero interband Josephson coupling.
\end{abstract}

\maketitle

\section{Introduction}
The majority of the superconductors of current interest have multiple superconducting
bands \cite{PhysRevLett.3.552,moskalenko1959superconductivity} with a widely varying strength of the interband coupling.
Understanding the boundary effects of a superconductor is
important since, superconducting currents are concentrated near the surfaces, and the physics of boundaries  controls vortex entry barriers and thus the onset of dissipation. Moreover, the behavior of the gaps near boundaries is crucial in 
small superconducting devices, such as superconducting nanowires and single-photon detectors, where multiband materials are  utilized
\cite{shibata2013fabrication}. 

In the last decades, topological superconductors have attracted particular interest. These materials exhibit topological surface currents, the observation of which is searched as a smoking gun for topological superconductivity and  can potentially be used to understand the nature of it. Among the candidate materials, there are compounds with complicated  multiband structure, raising the need of an understanding of the surface's properties \cite{bouhon2014current}.
For conventional and exotic multiband materials, the gaps are characterized by variety of probes, some of which 
selectively probe surfaces, while others are dominated by the bulk response
\cite{kreisel2020remarkable,chubukov2012pairing,mazin2003electronic,mackenzie2017even,sharma2020momentum,boker2017s+,zhao2020observation,suderow}.
The gap ratio is a characteristic quantity that allows one to get an insight into physics of Cooper pairing.
In the presence of surface superconductivity, it therefore important to study the gap ratio properties both in the bulk and near the boundaries.

The series of experimental works \cite{tsindlekht2004tunneling,belogolovskii2010zirconium,khasanov2005anomalous}
reported that, on the surface of  $\textrm{ZrB}_{12}$,
the characteristics of the superconducting gaps are widely different compared to the bulk.
Refs. \cite{gasparov2006two,biswas2020coexistence} suggested that $\textrm{ZrB}_{12}$ is a multiband superconductor with weak interband coupling.
A partial summary of the experimentally observed discrepancies, concerning surface/bulk gap structure, can be found in Table I in \cite{belogolovskii2010zirconium}. 
The surface effects are quite strong compared to  other reported experimental examples of enhanced surface superconductivity   \cite{fink1969surface,lortz2006origin, janod1993split, butera1988high,khlyustikov2011critical,khlyustikov2016surface,mangel2020stiffnessometer}.  
To explain this, it was proposed to search for
a mechanism of different phonon-electron interaction on the surface of the material \cite{ginzburg1964surface}.
However, the recent works
\cite{samoilenka2020boundary,samoilenka2020microscopic} reported that enhanced superconductivity near the boundary is a generic property of the standard single-band Bardeen-Cooper-Schrieffer model.
Namely, it was found that the presence of Friedel oscillations near the boundary induces an  increase  in the density of states,
at a microscopic length scale, yielding a higher critical temperature \cite{samoilenka2020microscopic}. The solution has multiple length scales and depends on the coherence length.

That raises the question of the nature of surface states in a generic multi-band Bardeen-Cooper-Schrieffer model \cite{PhysRevLett.3.552,moskalenko1959superconductivity}, where 
multiple length- and energy-scales are present.
In this work, we analyze the behaviour of the two superconducting gaps as a function of the interband coupling.
We focus on the  limit of a clean ideal surface, with negligible single-particle
interband scattering.
\footnote{The effects of single-particle interband scattering were already studied in \cite{bascones2001surface}. The effects of atomic-scale surface imperfections were addressed in \cite{samoilenka2020boundary} for a single-band material.}
 
\section{The model} 
We consider a Fermi-Hubbard Hamiltonian   describing a two-band $s$-wave superconductor. For a $d$ dimensional hypercubic lattice it reads
\begin{multline}
        H = \sum_{i,j,\sigma,\alpha} \psi^\dagger_{i\sigma\alpha}h_{ij\sigma\alpha}\psi_{j\sigma\alpha}\\
        - \sum_{i\alpha,\beta} V_{\alpha\beta} \psi^\dagger_{i\uparrow\alpha}\psi^\dagger_{i\downarrow\alpha}\psi_{i\downarrow\beta}\psi_{i\uparrow\beta}\, . 
\end{multline}
The roman indices $i,j$ label the position on a lattice with $N$ lattice points.
$\sigma=\uparrow,\downarrow$ indicates the spin, while $\alpha,\beta=1,2$ label the component.
Then $h_{ij\sigma\alpha} = -\mu \delta_{ij} - t\delta_{\abs{i-j},1}$, where $\abs{i-j} = 1$ if $i$ and $j$ are neighboring points in hypercubic lattice. $\mu$ is the chemical potential and $t$ the hopping coefficient.
In order to ensure the Hamiltonian to be hermitian, we have $h_{ij\sigma\alpha} = h^*_{ji\sigma\alpha}$ and $V_{\alpha\beta}=V^*_{\beta\alpha}$.
Then, following the steps in \cite{samoilenka2020pair}, we perform the mean field approximation. Introducing the Nambu spinors 
\begin{equation}
\begin{split}
    \Psi_{\alpha}^\dagger &= \mqty(\psi_{1\uparrow\alpha}^\dagger,\cdots,\psi_{N\uparrow\alpha}^\dagger,\psi_{1\downarrow\alpha},\cdots,\psi_{N\downarrow\alpha} )\\ 
    \Psi_{\alpha} &= \mqty(\psi_{1\uparrow\alpha},\cdots, \psi_{N\uparrow\alpha}, \psi_{1\downarrow\alpha}^\dagger,\cdots,\psi_{N\downarrow\alpha}^\dagger )^T\, ,
\end{split}
\end{equation}
the total mean field Hamiltonian reads
\begin{equation}
H_{MF} = \sum_{\alpha=1}^2 \Psi_{\alpha}^\dagger H_\alpha \Psi_{\alpha}\, .
\end{equation}
$H_\alpha$ is the $\alpha$-band Hamiltonian, defined as
\begin{equation}
H_\alpha = \mqty( h_{\uparrow \alpha} && \Delta_\alpha \\ \Delta_\alpha^\dagger && -h_{\downarrow \alpha}^T ) \, ,
\end{equation}
where the elements $h_{ij\sigma\alpha}$ have been defined above. Finally, the self-consistency equations for the gaps are:
\begin{equation}
\Delta_{i\alpha} = \sum_{\beta=1}^2V_{\alpha\beta}\langle c_{i\uparrow\beta}c_{i\downarrow\beta}\rangle_\beta \, .
\end{equation}
The thermal average $\langle \cdot \rangle_\beta$ means it is performed over the eigenvalues of the hamiltonian $H_\beta$. We can rewrite the self consistency equation by introducing the auxiliary vectors $(\vb{e}_i)_j = \delta_{i,j}$ and $(\vb{h}_i)_j = \delta_{j,i+N}$ as
\begin{equation}\label{self_cons_eq}
\Delta_{i\alpha} = - \sum_{\beta=1}^2V_{\alpha\beta}\vb{e}_i f(H_\beta)\vb{h}_{i} \, ,
\end{equation}
with $f(x) = \qty(1+e^{x/T})^{-1}$ being the Fermi-Dirac function.
We solve self-consistently for the gaps $\Delta_{i\alpha}$, using Chebyshev polynomial expansion method \cite{kernelPoly,chebCovaci,chebNagai}, with polynomial up to order 1000.
The convergence criterion we adopt is $|\Delta_{i\alpha}^{(n+1)} - \Delta_{i\alpha}^{(n)}|/|\Delta_{i\alpha}^{(n)}|\leq10^{-8}$, where $n$ numbers the iteration.
We consider both a 1D lattice with $N=1000$ sites and a 2D square lattice with $N_x N_y = 60\times60$.
The  solver is a custom CUDA implementation.
To calculate the critical temperatures, we solve the linearized version of the self consistency equation \eqref{self_cons_eq}.
For details, see \cite{samoilenka2020boundary}.

\section{Results}
\subsection{Effects of interband coupling}
\begin{figure}[ht]
    \centering
    \includegraphics[width=1.0\columnwidth]{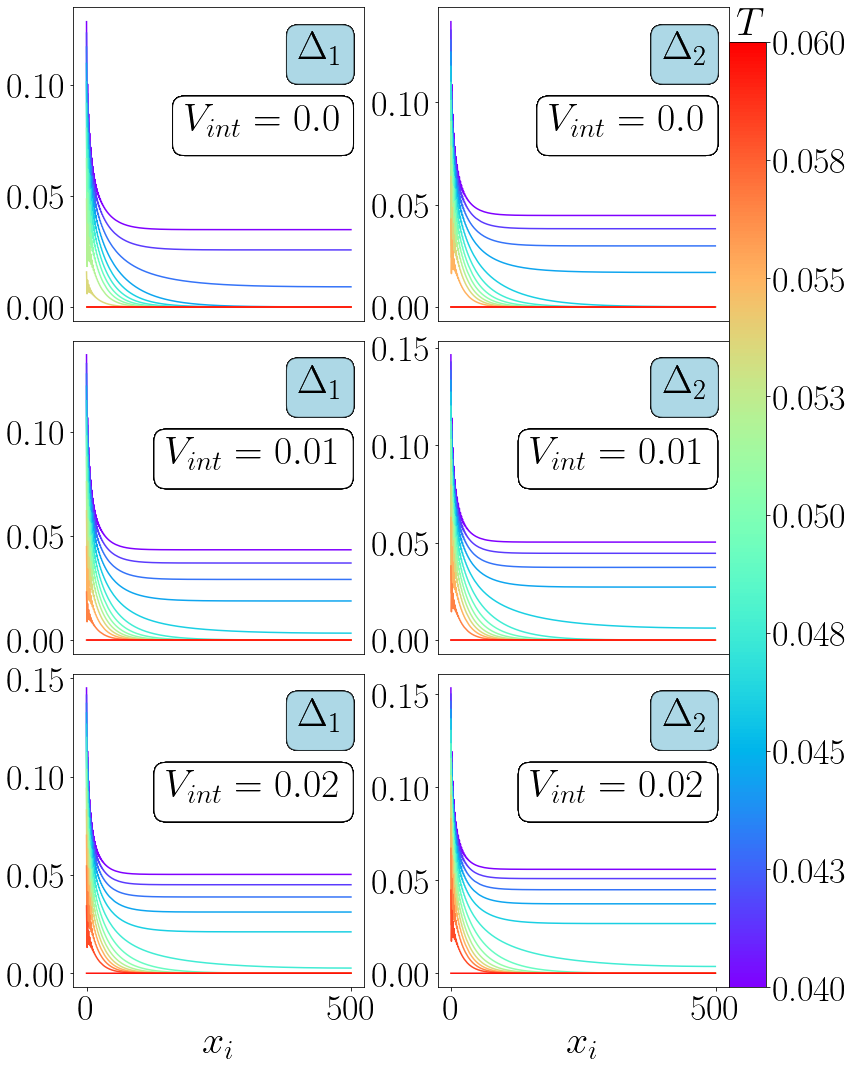}
    \caption{Numerical solution for the two gaps at different values of the interband coupling. The boundary is located at $x = 0$. For $V_{int} = 0.0$, the two bands have different boundary and bulk critical temperatures. When the interband interaction is on, the two bands present the same critical temperatures. Yet, the weak interband coupling does not drastically affect the structure of the solution. The gaps exhibit different enhancement near the boundary and the overall solution shows the presence of different length scales.
    For $V_{12}=0.01$ we have $T_{c1} = 0.0458$ and $T_{c2} = 0.0573$. For $V_{12}=0.02$ the critical temperatures increase respectively to $T_{c1} = 0.0472$, $T_{c2} = 0.0591$.
    Here we show only the left half of the system.}
    \label{fig:gaps}
\end{figure}

We begin by analyzing a 2-band system with weak interband coupling and similar intraband interaction in two  bands.
The model is rescaled so that all the quantities are given in units of the hopping coefficient $t$.
We fix $\mu=0$, i.e. half filled bands, $V_{11}=1.35$ and $V_{22}=1.36$. We compare the results for non-zero interband interaction $V_{12}$ with the case where the bands are decoupled, i.e. $V_{12}=0.0$. 
In the latter, the problem is reduced to two copies of the model studied in \cite{samoilenka2020boundary} and shown to have two different critical 
temperatures, one for bulk states and one for boundary states.
We denote by $T_{c1}$ the bulk critical temperature, i.e. when the order parameter vanishes in the bulk. $T_{c2}$ is the boundary critical temperature, i.e. when the order parameter vanishes on the boundaries of the superconductor.
When $V_{12}=0$ the critical temperatures in the system are: $T_{c1}^{\textrm{band1}}=0.0429$ and $T_{c2}^{\textrm{band1}}= 0.0536$ for band 1, $T_{c1}^{\textrm{band2}}=0.045$ and $T_{c2}^{\textrm{band2}}= 0.0562$ for band 2. 
Hence, in this example, the second band has a critical temperatures $5\%$ higher than the first.
Figure \ref{fig:gaps} shows the numerically obtained gaps $\Delta_1$ and $\Delta_2$, displayed at various temperatures and interband coupling. 
In accordance with the results obtained by different analytical and numerical methods in \cite{samoilenka2020boundary}, when $V_{12}=0$, the boundary-enhancement of each gap decays to the bulk value with independent coherence length.
As the coupling is turned on, $U(1)$ is broken, since the carriers in the individual bands are no longer independently conserved and there are no independent transitions for different bands. 
\begin{figure}[H]
    \centering
    \includegraphics[width=1\columnwidth]{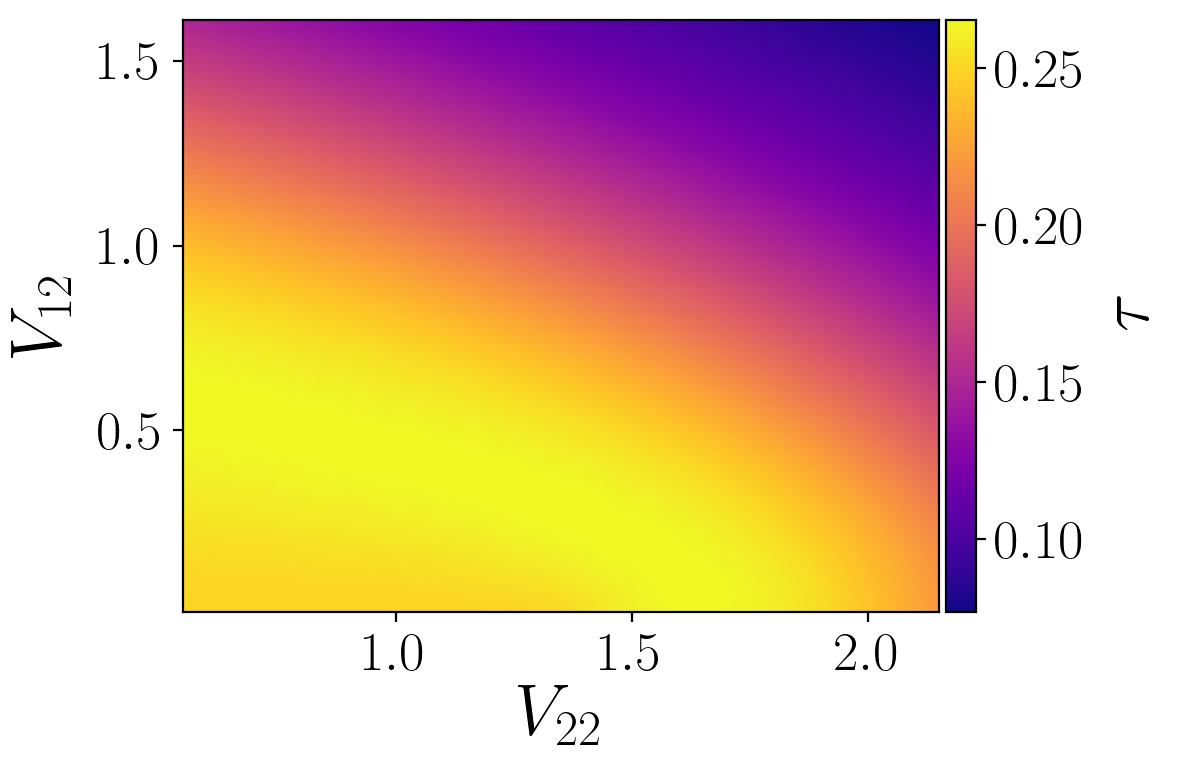}
    \caption{Relative increase of the boundary critical temperature with respect to the bulk critical temperature as a function of the pairing potential of the second band $V_{22}$ and of the inter-band coupling $V_{12}$. We define $\tau = \qty(T_{c2}-T_{c1})/T_{c1}$.
    For a given value of $V_{22}$, $\tau$ exhibit a non monotonic behavior as a function of interband coupling $V_{12}$. $V_{11}=1.35$ and $\mu=0$.}
    \label{fig:Tc2d}
\end{figure}
Then, for non-zero $V_{12}$, the bulk  critical temperatures become the same for both bands. Also the surface critical temperature is only one.
The gaps behaviour near the boundaries is nontrivial as it includes relative variations of the gap values.
In the case displayed in Figure \ref{fig:gaps} we have $T_{c1} = 0.0458$ and  $T_{c2} = 0.0573$ for $V_{12}=0.01$; $T_{c1} = 0.0472$ and $T_{c2} = 0.0591$ for $V_{12}=0.02$. 

We conclude this section by moving beyond the weak interband coupling regime and
investigate  the relative increase of the boundary critical temperature $T_{c2}$, with respect to the bulk temperature $T_{c1}$, as a function of inter-band coupling $V_{12}$.
To efficiently measure this increase, we define
$\tau = \qty(T_{c2}-T_{c1})/T_{c1}$.
The numerical solutions for a one dimensional model are shown on the Figure \ref{fig:Tc2d} for various values of $V_{22}$.  We find that the dependence is non-trivial: at relatively weak interband coupling, $\tau$ first increases with $V_{12}$ and then, it starts to decrease.

\onecolumngrid
\begin{figure*}[t]
    \includegraphics[width=1.9\columnwidth]{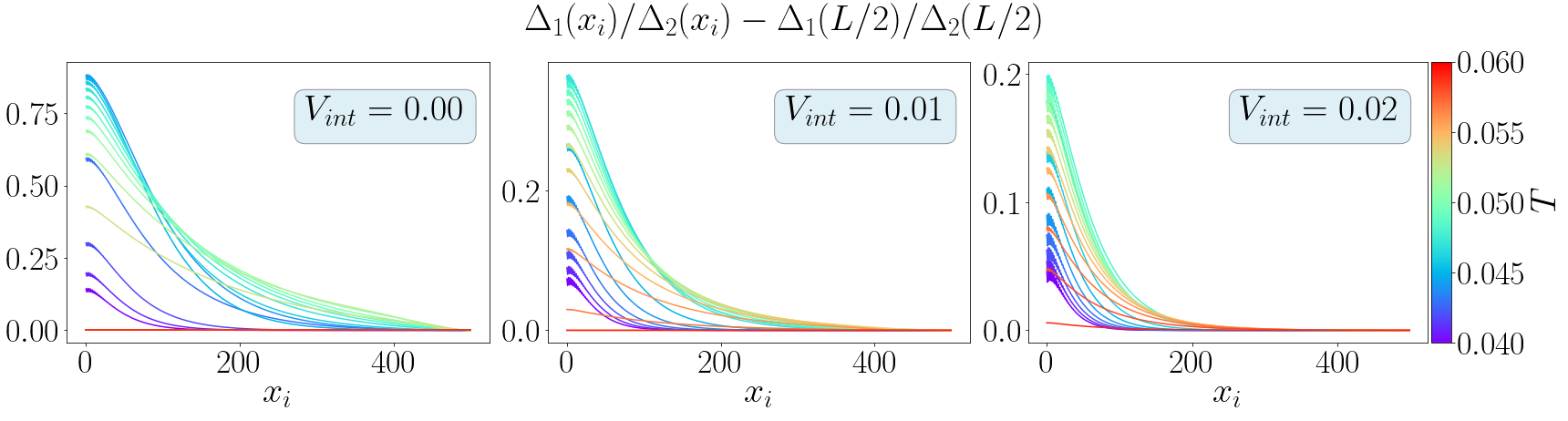}
    \caption{Plot of gap ratios shifted by the bulk value. We can notice that the gaps ratio near the surface is enhanced compared to the bulk value. The presence of weak inter-band coupling does not qualitatively change this effect. Moreover, the length scale of the penetration into the bulk shows a non monotonic behaviour as a function of $T$, as more accurately displayed in Figure \ref{fig:xi}. We show only half of the system ($500$ out of $L=1000$ sites), since the second half is entirely symmetrical. The parameters used for this simulations are $V_{11}=1.35$, $V_{22}=1.36$ and $\mu=0$.}
    \label{fig:gapRatio1}
\end{figure*}
\twocolumngrid
\subsection{The relative behavior of the gaps in two-band systems: boundaries vs bulk}
A useful characteristic of a multiband superconductor is the ratio of the gaps of different bands, whose temperature dependence can give insights into nature of pairing in the material.
Figure \ref{fig:gapRatio1} shows the gaps ratio shifted by its bulk values at various temperatures $T$ and inter-band coupling $V_{12}$ for a 1D system as a function of distance from the boundary. 
The system we consider first has the intraband potential of the second band only $1\%$ bigger than the first one, namely $V_{11}=1.35$ and $V_{22}=1.36$.

Even for these similar gap characteristics, we find that the gaps ratio can be different 
on the boundary of a superconductor
compared to its bulk value, when the interband coupling is weak. Figure \ref{fig:gapRatio1} displays the results for  $V_{12}=0.01$, $0.02$ and compares them to the decoupled-bands case.

\begin{figure}[ht]
    \centering
    \includegraphics[width=1.0\columnwidth]{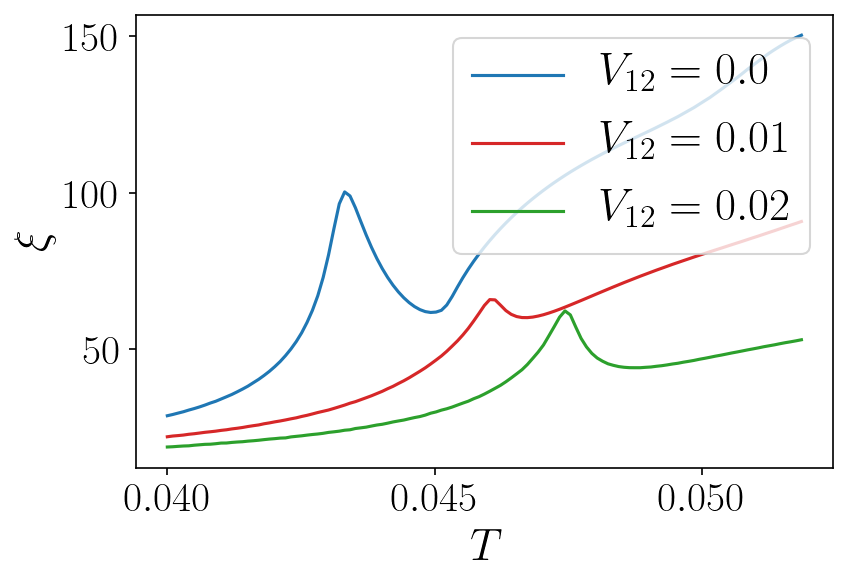}
    \caption{
    Long range asymptotic decay length scale $\xi(T)$ of the gaps ratio deviation displayed in Figure \ref{fig:gapRatio1}. $\xi(T)$ is plotted as a function of the temperature $T$ and for different values of the interband coupling. The non monotonic behavior as a function of $T$ is clearly visible. We obtain $\xi(T)$ by fitting the tails (i.e. after $N=50$ sites from the boundary) of the gaps ratio deviation using an exponential function $f(x) \propto e^{-x/\xi}$, for the different values of $T$ and $V_{12}$. The remaining parameters used in the simulations are $V_{11}=1.35$, $V_{22}=1.36$ and $\mu=0$.}
    \label{fig:xi}
\end{figure}

In particular, we notice that the both the gaps and their ratio are enhanced near the ends of the sample, and this enhancement decays into the interior of the superconductor on a macroscopic length scale. 
The surface gaps ratio deviation has not only a strong temperature dependence in magnitude, but also its length scale varies as a function of $T$. We can study the latter's behavior in further details by fitting the tails (i.e. after N=50 sites from the boundary) of the gaps ratio deviations reported in Figure \ref{fig:gapRatio1}, with an exponential function $f(x) \propto e^{-x/\xi}$. Here, $\xi(T)$ measures the length scale of the decay into the bulk. The result, reported in Figure \ref{fig:xi} confirms the non-monotonic behavior of $\xi(T)$ as a function of $T$. 

\begin{figure}[ht]
    \centering
    \includegraphics[width=0.8\columnwidth]{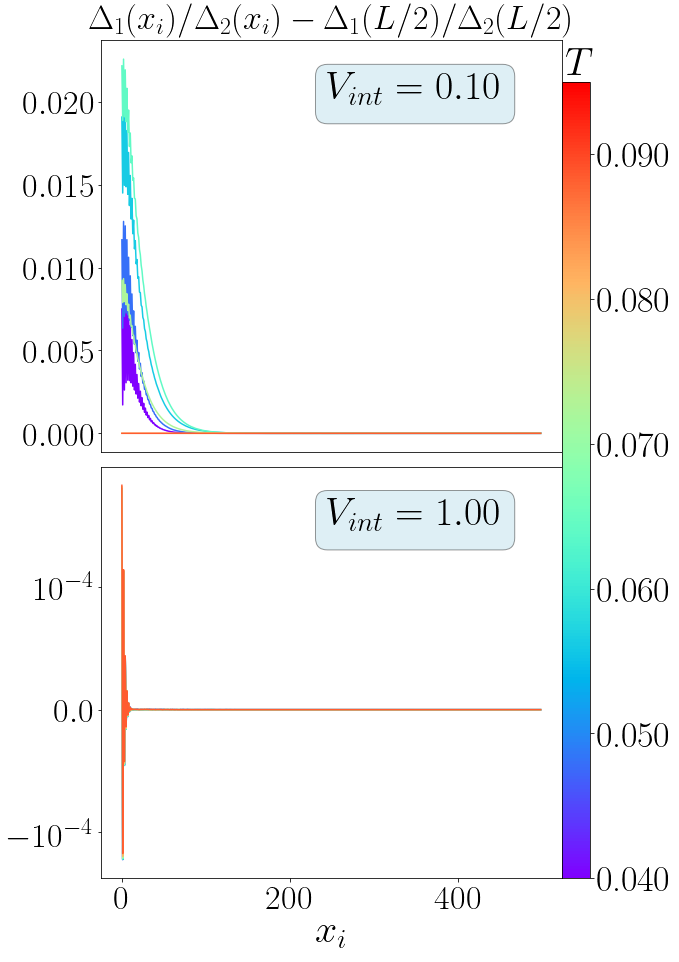}
    \caption{Suppression of the relative gap variations at stronger interband coupling. We can notice that the gaps ratio change near the boundary is an order of magnitude smaller for $V_{12}=0.1$ than for $V_{12}=0.01$. For strong interband coupling $V_{12}=1$, in the simplest two-band model,  the surface-induced change of the gaps ratio is  negligible. Also in this case we show only half of the system (500 out of $N$ = 1000 sites), since the second half is  symmetrical.The parameters used for these calculations are $V_{11}=1.35$, $V_{22}=1.36$ and $\mu=0$.}
    \label{fig:gapRatio2}
\end{figure}

Discussion concerning the existence of multiple length scales in multiband material, previously only focused 
on vortex physics \cite{Silaev2011,silaev2012microscopic,Carlstrom.Babaev.ea:11}.
The behavior of the surface states that we find is a new example of existence of multiple length scales in multiband materials, despite non-zero Josephson coupling. The long range character associated with the relative variations of the gaps, and its non-monotonicity is consistent with the conclusions obtained for the vortex core solutions in weakly interacting two-band systems in \cite{Silaev2011,silaev2012microscopic}.
At higher values of the inter-band coupling, we can notice that the relative
variation of the gaps profile near the surfaces 
decreases, both in amplitude and in its spatial extension, as Figure \ref{fig:gapRatio2} reports. This remains consistent with the hybridization of bulk coherence lengths and their dependence on interband coupling strength \cite{Silaev2011}.
When the interband coupling $V_{12}$ becomes of the same order of magnitude as the intraband coupling $V_{11}$ and $V_{22}$, e.g. $V_{12}=1.0$, the enhancement of the gaps ratio basically disappears. Note that the disappearance of this variation is similar to the condition for the disappearance of the second
coherence lengths in the clean two-band BCS semi-classical model found in \cite{Silaev2011}. This confirms that the width of the boundary states, in two band states, is in general determined by two bulk coherence lengths.

\begin{figure}[t]
    \centering
    \includegraphics[width=1\columnwidth]{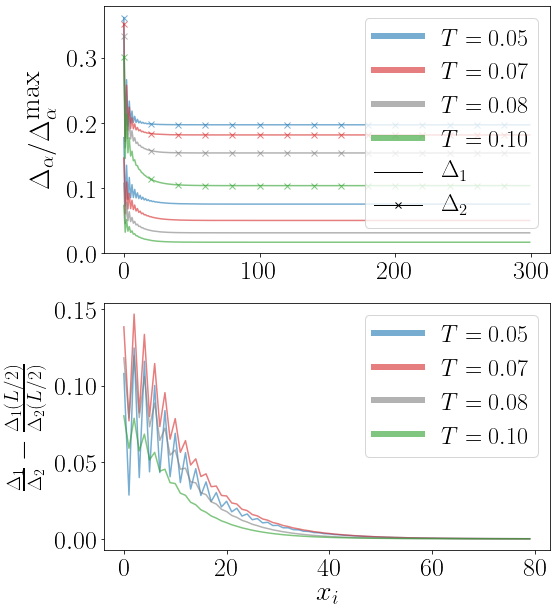}
    \caption{Numerically obtained order parameters for different values of the $T$. In this case, we study a coupled two-band system where the difference between the two intraband potentials $V_{11}$ and $V_{22}$ is higher, namely $25\%$. The upper panel displays the gaps $\Delta_1$ (solid line) and $\Delta_2$ (crossed line) independently.
    The bottom panel reports the gaps ratio near-surface spatial variation relative to the bulk value.
    For both panels we fixed $V_{12}=0.1$. We can notice that the modest increase in $V_{22}$ results
    in substantially larger variation of the
    gap ratio  near the boundary.
    Both results were obtained for a system with $N=1000$ and $\mu=0$.} 
    \label{fig:higerV22}
\end{figure}

Next we consider the boundary states when the difference between the intraband potential is greater. Specifically we consider  $V_{11}=1.35$ and $V_{22}=1.68$. 
The upper panel of Figure \ref{fig:higerV22} shows the gaps $\Delta_1$ (solid line) and $\Delta_2$ (dashed line) 
at various temperatures. The bottom panel displays the variation of the gaps ratio with respect to the bulk value. Here the interband potential is set to be $V_{12}=0.1$. We can see a moderate increase of $V_{22}$ yields a substantial
variation of the relative gap  values near the surface compared to Figure \ref{fig:gapRatio2}.
\subsection{Surface effects in two-dimensions and corner states}
\begin{figure*}[ht]
    \centering
    \includegraphics[width=1.8\columnwidth]{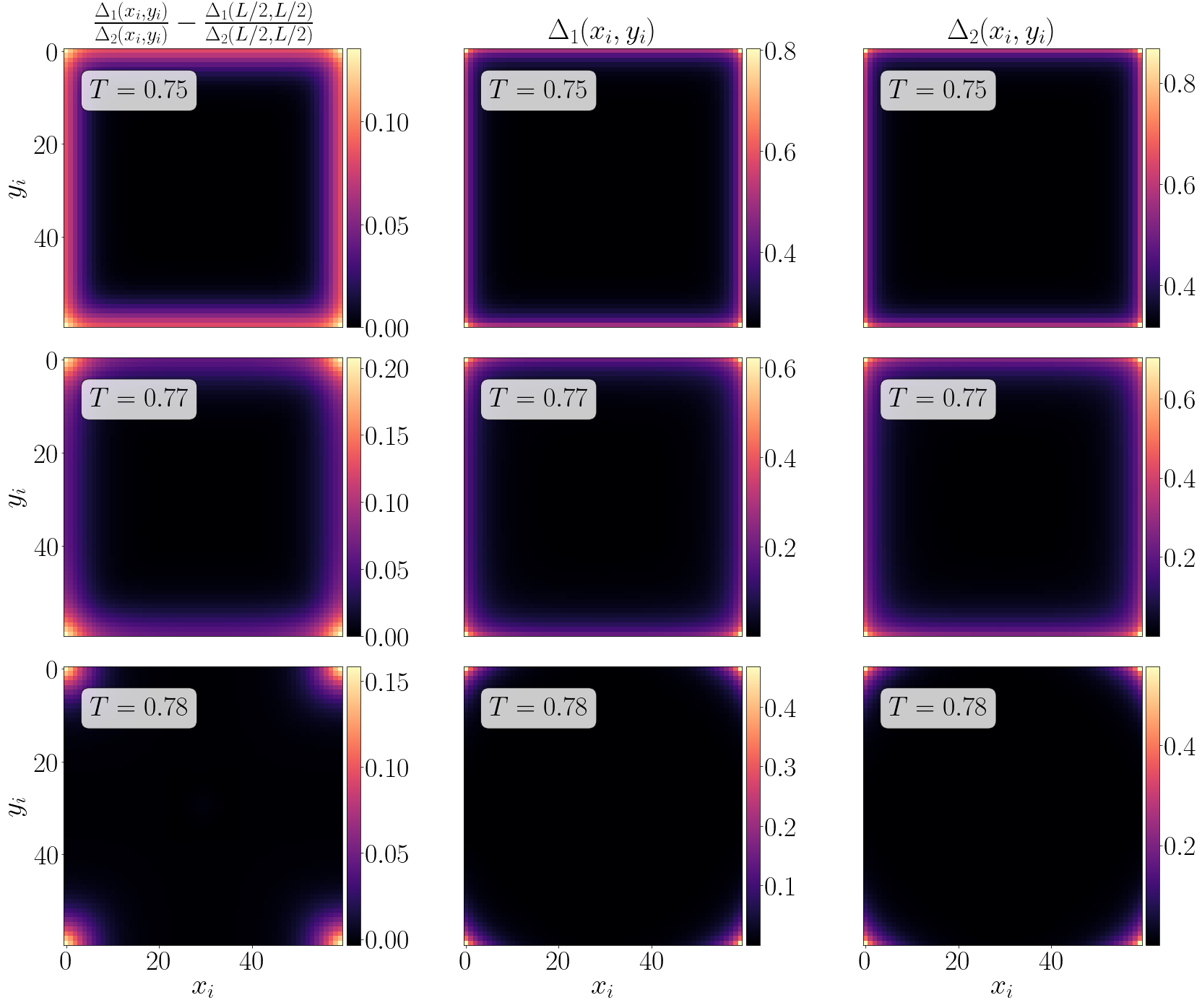}
    \caption{Plot of the gaps ratio shifted by the bulk value (left column), $\Delta_1$ (center column) and $\Delta_2$ (right column) in two dimensions for increasing temperatures. The bulk critical temperature for the system reads $T_{c1} = 0.759$ and the edge critical temperature is $T_{c2} = 0.774$. Therefore the first row shows the bulk superconductivity, the second raw shows the edge superconductivity and the third row shows the state where the gap survives only in the corners. Below the corner critical temperature, i.e. for bulk and edge states we can notice the gradient  of the gaps ration localized along the system's boundaries. The increase is more pronounced above the bulk critical temperature, i.e. with edge states. When $T_{c2}<T<T_{c3}$ superconductivity survives in the four corners, where also the gaps ratio undergoes significant enhancement, penetrating into the bulk with macroscopic length scale.}
    \label{fig:2d}
\end{figure*}

In 2D and 3D single-component BCS models there are superconducting corner and edge states with relative critical temperature higher than 
the bulk critical temperature \cite{samoilenka2020pair,samoilenka2020boundary}. 
In this section, we consider the  gaps ratio spatial profile  in a two dimensional two-band system.
In 2D we have edges and corners, therefore we can associate $T_{c2}$ as the mean-field critical temperature for edge superconductivity, and $T_{c3}$ as the mean-field  critical temperature for corner superconductivity.
In a single-band BCS 2D system, $T_{c3}>T_{c2}$, as shown in \cite{samoilenka2020pair}.
We study the two-band system for $T<T_{c1}$, $T_{c1}<T<T_{c2}$ and $T_{c2}<T<T_{c3}$.
Studying the boundary effects in two dimensions is challenging, as it requires numerically solving significantly large systems, to avoid the finite-size effects' influence on the resulting states. 
Figure \ref{fig:2d} shows the gaps ratio shifted by its value in the bulk, and the two gaps $\Delta_1$, $\Delta_2$.
We can notice that the boundary states exist at much smaller length scales than the size of the sample.
Both for the bulk ($T=0.75$) and the edge ($T=0.77$) states the gaps ratio is enhanced along the system boundaries. 
When the temperature exceeds the edge critical temperature $T_{c2}=0.774$, the superconductivity ceases to exists along the boundaries, but remains in the four corners. We observe that in the corners there is the
largest variation of the gaps ratio, which decays into the bulk at  a macroscopic length scale.

\section{Conclusions}

Since most of superconductors of current interest are
multiband, and there are experimental
puzzles, such as the surface gap enhancement   in $\textrm{ZrB}_{12}$ \cite{tsindlekht2004tunneling,belogolovskii2010zirconium,khasanov2005anomalous} it is important to understand the boundary effects in multiband superconducting materials.

We studied the boundary effects in the standard two-band Bardeen-Cooper-Schrieffer theory of superconductivity.
We showed that, at the level of mean-field theory, the system has multiple critical temperatures, associated to the presence of boundary states.

We found that the dependence of the critical temperatures on the value of interband coupling is non-monotonic.

Moreover, when   interband coupling is relatively
weak, the behavior of the gaps near the boundaries presents multiple coherence lengths and a relative variation of the gaps values.

In dimensions higher than one, the effects are stronger in the sample's corners. The relative variation of the gaps values extends into the superconductor with a large temperature-dependent length scale.
This should be of particular importance for critical currents in small superconducting devices. 
The enhanced surface superconductivity
may be harvested to improve superconducting-nanowire-based single photon detectors. A material with increased gap near the surface is expected to enhance vortex entry barrier which can yield less dark counts.
An interesting question for further studies is the effect of boundary-induced single particle scattering \cite{bascones2001surface} on the states we report.

\section{Acknowledgements}
The work was supported by the Swedish Research Council Grants No. 642-2013-7837, 2016-06122, 2018-03659, the G\"{o}ran Gustafsson Foundation for Research in Natural Sciences and Medicine, Olle Engkvists Stiftelse.
\bibliography{bibliography.bib}
\end{document}